\documentclass[prl,aps,twocolumn,showpacs,superscriptaddress]{revtex4-1}
\usepackage{hyperref}
\hypersetup{colorlinks=true, citecolor=blue, filecolor= black, linkcolor= blue, urlcolor= blue}
\usepackage{graphicx}% Include figure files
\usepackage{dcolumn}% Align table columns on decimal point
\usepackage{bm}% bold math
\begin{document}
%\begin{CJK*}{GB}{kai}
%\CJKfamily{gbsn}

%\preprint{Submitted to PRL}

\title{Unexpected Reconstruction of the $\alpha$-Boron $\mathbf{(111)}$ Surface}

\author{Xiang-Feng Zhou}
\email{xfzhou@nankai.edu.cn}
\email{zxf888@163.com}
\affiliation{School of Physics and Key Laboratory of Weak-Light Nonlinear Photonics, Nankai University, Tianjin 300071, China}
\affiliation{Department of Geosciences, Center for Materials by Design, and Institute for Advanced Computational Science, Stony Brook University, Stony Brook, New York 11794, USA}
%\thanks{}
\author{Artem R. Oganov}
\affiliation{Department of Geosciences, Center for Materials by Design, and Institute for Advanced Computational Science, Stony Brook University, Stony Brook, New York 11794, USA}
\affiliation{Moscow Institute of Physics and Technology, 9 Institutskiy lane, Dolgoprudny city, Moscow Region 141700, Russian Federation}
\affiliation{School of Materials Science, Northwestern Polytechnical University, Xi'an 710072, China}

\author{Xi Shao}
\affiliation{School of Physics and Key Laboratory of Weak-Light Nonlinear Photonics, Nankai University, Tianjin 300071, China}

\author{Qiang Zhu}
\affiliation{Department of Geosciences, Center for Materials by Design, and Institute for Advanced Computational Science, Stony Brook University, Stony Brook, New York 11794, USA}

\author{Hui-Tian Wang}
\affiliation{School of Physics and Key Laboratory of Weak-Light Nonlinear Photonics, Nankai University, Tianjin 300071, China}
\affiliation{National Laboratory of Solid State Microstructures, Nanjing University, Nanjing 210093, China}
%\date{\today}

\begin{abstract}
\noindent We report on a novel reconstruction of the $\alpha$-boron $(111)$ surface, discovered using an \textit{ab initio} evolution structure search, and reveal that it has an unexpected neat structure and much lower surface energy than the recently proposed $(111)$-I$_{R,(a)}$ surface. For this reconstruction, every single interstitial boron atom forms bridges with the unique polar-covalent bonds between neighboring $B_{12}$ icosahedra, which perfectly meet the electron counting rule and are responsible for the reconstruction-induced metal-semiconductor transition. The peculiar charge transfer between the interstitial atoms and the icosahedra plays an important role in stabilizing the surface.
\end{abstract}

\pacs{68.35.bg, 71.15.Mb, 73.20.At}

%\keywords{}

\maketitle
%\end{CJK*}
%\newpage
%\begin{multicols}
\noindent The element boron has been attracting an enormous amount of attention owing to its fascinating properties, such as fascinating structural complexity, superhardness, unusual partially ionic bonding, and superconductivity at high pressure \cite{R01,R02,R03,R04,R05}. As a neighbor for carbon, boron is in many ways an analog of carbon and its nanostructures (clusters, nanotubes, nanowires, nanobelts, fullerenes and so on) have aroused extensive interest, in the hopes also replicating or even surpassing the unique properties and diversity of carbon \cite{R06,R07,R08,R09,R10,R11,R12}. In analogy to graphene \cite{R13,R14}, two dimensional (2D) boron sheets with the triangular and hexagonal motifs are predicted to be the most stable phases and likely precursors for boron nanostructures \cite{R15,R16,R17,R18,R19,R20,R21,R22}. However, buckled bilayer structures appeared to be massively more stable; some of them turned out to have novel electronic properties, such as a distorted Dirac cone \cite{R23}. Surprisingly, there are a few studies on the boron surface with the dimension between bulk and 2D sheets. Hayami and Otani systematically studied the energies of low index bare surfaces in the $\alpha$-boron, $\beta$-boron, and two tetragonal phases (t-I and t-II), which suggested that t-I and t-II can be more stable than $\alpha$-boron and $\beta$-boron for sufficiently small nanoparticles \cite{R24,R25}. Amsler \textit{et al}. took the first big step on the reconstruction of the $\alpha$-boron $(111)$ surface. Several low energy surface reconstructions were predicted by using the minima hopping method. In particular, a metallic reconstructed phase of $(111)$-I$_{R,(a)}$ was predicted to be the most stable configuration, where a conducting boron sheet was adsorbed on a semiconducting substrate, leading to numerous possible applications in nanoelectronics \cite{R26}. However, this seems to be in conflict with the general principle that the reconstructions usually lower their energies by atomic rearrangement leading to semiconducting (as opposed to metallic) surface state \cite{R27}. Such an unexpected metallic reconstruction encourages us to explore other likely reconstructions and the stabilization mechanisms by first-principles calculations.

$\alpha$-boron structure is composed of $B_{12}$ icosahedra \cite{R28}, has two inequivalent atomic sites, polar ($B_{p}$) and equatorial ($B_{e}$) sites, the $B_{p}$ atoms form upper and lower triangles of an icosahedron and the $B_{e}$ atoms form a waving hexagon along the equator of an icosahedron \cite{R02}. The arrangement of icosahedrz in $\alpha$-boron can be described as a cubic close packing with the layer sequence ABC \cite{R01}. Compared with the $(111)$-I$_{R,(a)}$ surface that built along $[1 1 1]$ direction of the primitive rhombohedral cell with the surface vectors $\mathbf{U} (2 \bar{1} \bar{1})$ and $\mathbf{V} (1 1 \bar{2})$, we cleaved the surface along $[1 1 1]$ direction with the surface vectors $\mathbf{U} (\bar{1}  \bar{1} 2)$ and $\mathbf{V} (1 \bar{1} 0)$, which allowed us to reduce the required computational resource drastically. The calculations were conducted on the 4 layered $B_{12}$ icosahedra of the $(111)$-I substrate, and then $(111)$-II substrate was also tested \cite{R25,R26}. In both cases we obtained exactly the same reconstruction, regardless the type of substrate if enough atoms and thickness are used. Structure searches for the reconstructions were performed using the \textit{ab initio} evolutionary algorithm \textsc{uspex} \cite{R29,R30,R31}, which has been successfully applied to various materials \cite{R32,R33,R34}. The number of surface atoms was allowed to vary from 1 to 20 with the vacuum layer of 10 {\AA}, which are restricted to the surface layer of thickness 4~{\AA}. Given that the thickness of $B_{12}$ icosahedron is $\sim$3.7~{\AA}, there is an enough space to fully explore the chemical landscape in our calculations. The all-electron projector-augmented wave method \cite{R35} was employed, as implemented in the Vienna \textit{ab initio} simulation package (VASP) \cite{R36} with the generalized gradient approximation (GGA) and the functional of Perdew, Burke, and Ernzerhof (PBE) \cite{R37}. A plane-wave cutoff energy of 500~eV and a Monkhorst-Pack Brillouin zone sampling grid with resolution $2 \pi \times 0.04$~\AA$^{-1}$ were used. In addition, the hybrid HSE06 functional with the screening parameter ($\omega$) 0.2~{\AA}$^{-1}$ was also employed to check the robustness of surface energies \cite{R38}.

\begin{figure}[h]
\begin{center}
\includegraphics[width=8.5cm]{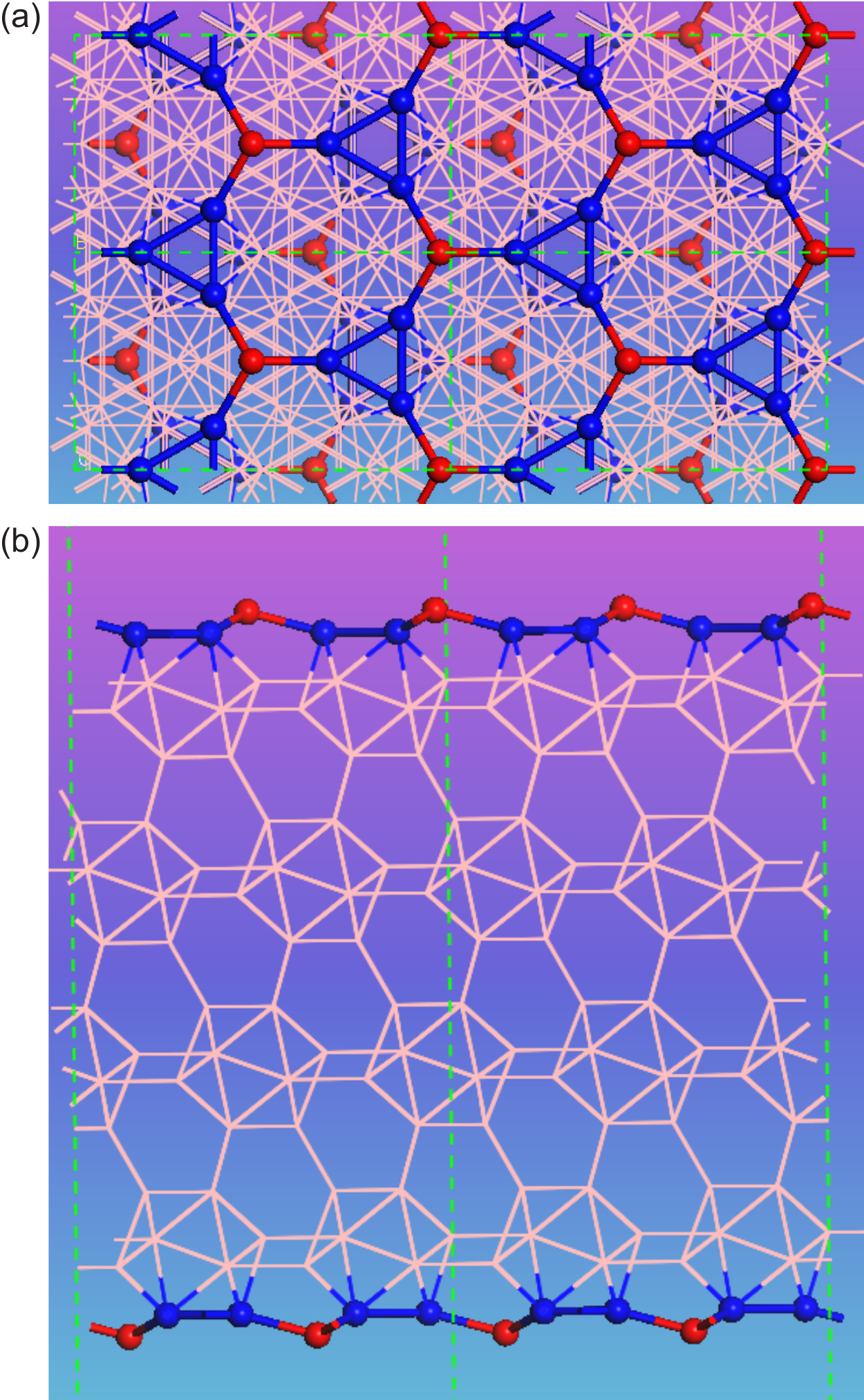}
\caption{%
(Color online) (a) Projection of $2 \times 2 \times 1$ supercell of the $(111)$-I$_{R,(z)}$ structure along $[111]$ direction. (b) projection of $2 \times 2 \times 1$ supercell of the $(111)$-I$_{R,(z)}$ structure along $[\bar{1} \bar{1} 2]$ direction. The nonequivalent surface atoms are shown by different colors.}
%\label{fig:eos}
\end{center}
\end{figure}

For the most stable $(111)$-I$_{R,(a)}$ surface, there is buckling and coupling among three outer atomic layers above the icosahedral $B_{12}$ units, which result in structural complexity \cite{R26}. In contrast to $(111)$-I$_{R,(a)}$, our reconstruction [designated as $(111)$-I$_{R,(z)}$] has an unexpected neat structure, as shown in Fig.~1(a), where a single boron atom occupies the interstitial position (named as $B_{i}$, colored in red), connects the $B_{p}$ atoms (colored in blue) and forms bridges with bond lengths of 1.793 {\AA} and bond angles of $113.06^\circ$. The $B_{i}$ atoms are slightly above the topmost icosahedral atoms ($B_{p}$), see Fig.~1(b), which form the modulated ``$3+9$" membered rings on the topmost surface. Interestingly, the $(111)$-I$_{R,(z)}$ and $(111)$-I$_{R,(a)}$ share almost the same surface motif ($3+9$ structure), which hints that the $(111)$-I$_{R,(a)}$ surface may be a local metastable phase. According to the surface energy $\sigma$ = (1/2$A$) ($N$$\epsilon_{bulk}$ - $E_{slab}$), where $A$ denotes the surface area, $\epsilon_{bulk}$ is the energy per atom in the bulk $\alpha$-boron and $E_{slab}$ is the energy of the substrate containing $N$ atoms \cite{R26}. We calculate the surface energies of $(111)$-I and $(111)$-I$_{R,(a)}$ by using the GGA-PBE and HSE06 methods, as listed in the Table I, which are in excellent agreement with previous results \cite{R26}. This establishes the reliability and accuracy of our calculations. Strikingly, the surface energy of $(111)$-I$_{R,(z)}$ is 128.23 meV/{\AA}$^{2}$ for GGA-PBE and 136.79 meV/{\AA}$^{2}$ for HSE06, which is considerably, by 42 and 60 meV/{\AA}$^{2}$ respectively, lower in energy than the $(111)$-I$_{R,(a)}$ structure. To confirm the most stable surface, we also perform the structure search with the same substrate proposed by Ref. [26], get the same reconstruction and energy, and find there is no dependence on the choice of the surface vectors/cleavage plane: it is the general rule that the ratio of $B_{i}$ to the exposed $B_{12}$ icosahedron should be $1:1$ in $\alpha$-boron $(111)$ surface.

\begin{table}
\caption{The surface energies of the unreconstructed $(111)$-I, reconstructed $(111)$-I$_{R,(a)}$, and $(111)$-I$_{R,(z)}$ structures by using different functionals, in units of meV/{\AA}$^{2}$.}
\begin{tabular}{llllc}
\hline\hline
Surface &$(111)$-I &$(111)$-I$_{R,(a)}$ &$(111)$-I$_{R,(z)}$ & Reference \\
\hline
PBE & 219.29 & 170.64 & 128.23 & This work \\
    & 218.80 & 170.61 &        &   26    \\
HSE & 248.43 & 197.39 & 136.79 & This work \\
    & 247.50 & 196.31 &        &   26    \\
\hline\hline
\end{tabular}
\end{table}

Figure~2 shows the band structures of the unreconstructed $(111)$-I and the reconstructed $(111)$-I$_{R,(z)}$ surfaces from the GGA-PBE calculations. Due to the unsaturated dangling bonds, the $(111)$-I surface is metallic, as shown in Fig.~2(a). In contrast, Figure 2(b) shows the $(111)$-I$_{R,(z)}$ surface is a semiconductor with a direct DFT band gap of 1.13~eV compared with the band gap of 1.50~eV for bulk $\alpha$-boron. The reconstruction-induced metal-semiconductor transition meets the general principle that reconstructions usually lower their energies by atomic rearrangements leading to semiconducting surface state \cite{R27}. According to Wade's rule \cite{R39,R40}, a $B_{12}$ icosahedron has 36 valence electrons, 26 of which may be used for intraicosahedral bonds and 10 for intericosahedral bonds. Each icosahedron forms six two-electron-two-center (2e2c) bonds with the icosahedra of neighboring layers, which requires $6 \times 2/2 = 6$ electrons, as well as six closed two-electron-three-center (2e3c) bonds with the neighboring icosahedra in its own layer, these multicenter bonds require $6 \times 2/3 = 4$ electrons \cite{R01}. The $(111)$-I surface cuts three intericosahedral bonds (2e2c) per icosahedron, and there are two $B_{12}$ icosahedra per stacking layer in the $(111)$-I$_{R,(z)}$ surface. Therefore, the unsaturated dangling bonds of the $(111)$-I$_{R,(z)}$ surface need additional $3 \times 2 \times 2/2 = 6$ valence electrons. Two $B_{i}$ atoms are just added to the surface, connect six $B_{p}$ atoms, as shown in Fig.~1(a), form six bridging B-B bonds (these are the closed 2e2c bonds, may not the same), which perfectly satisfy the electron counting rule (ECR) \cite{R27,R41}, and are responsible for the reconstruction-induced metal-semiconductor transition.

\begin{figure}[h]
\begin{center}
\includegraphics[width=8.5cm]{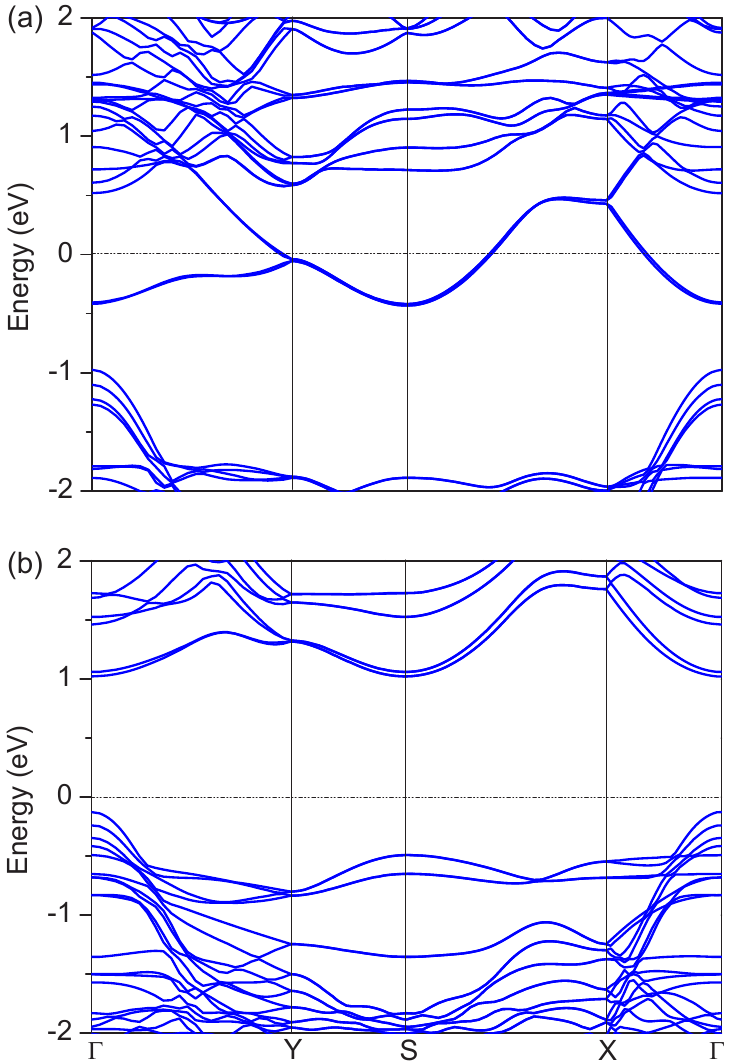}
\caption{%
(Color online) Band structures of (a) $(111)$-I and (b) $(111)$-I$_{R,(z)}$, the special $k$ points are labeled as $\Gamma$(0 0 0), $Y$(0 0.5 0), $S$(--0.5 0.5 0), and $X$(--0.5 0 0), respectively.}
%\label{fig:eos}
\end{center}
\end{figure}

Projected density of states (PDOS) of the topmost atoms is plotted in Fig.~3. The $(111)$-I surface exhibits metallic character, see Fig.~3(a), dominantly due to the out-of-plane states ($p_z$ orbitals), arising from the unsaturated $B_{p}$ atoms, and located near the bottom of the conduction band. In comparison, strongly hybridized bonding states present in the vicinity of the Fermi level in Fig.~3(b), mainly derive from the $B_{p}$: $p_z$ and $B_{i}$: $p_{xy}$ orbitals, and are fully filled. The $(111)$-I$_{R,(z)}$ surface is thereby a semiconductor. PDOS (Figs. 3c and 3d) clearly shows the out-of-plane $p_z$ states and the in-plane $p_{xy}$ states near the valence band edge are dominantly from the $B_{p}$ and $B_{i}$ atoms, respectively. Because the $B_{i}$ atoms are located above empty space, there is no $p_z$ state for the $B_{i}$ under the Fermi level in Fig.~3(d). Moreover, the distance between the $B_{i}$ and $B_{e}$ atoms is 2.926~{\AA}. All of these facts indicate that there is no interaction/bonding between the $B_{i}$ and $B_{e}$ atoms, which further confirms the reliability of the surface bonding configuration and the ECR applied for the $(111)$-I$_{R,(z)}$ surface.

\begin{figure}[h]
\begin{center}
\includegraphics[width=8cm]{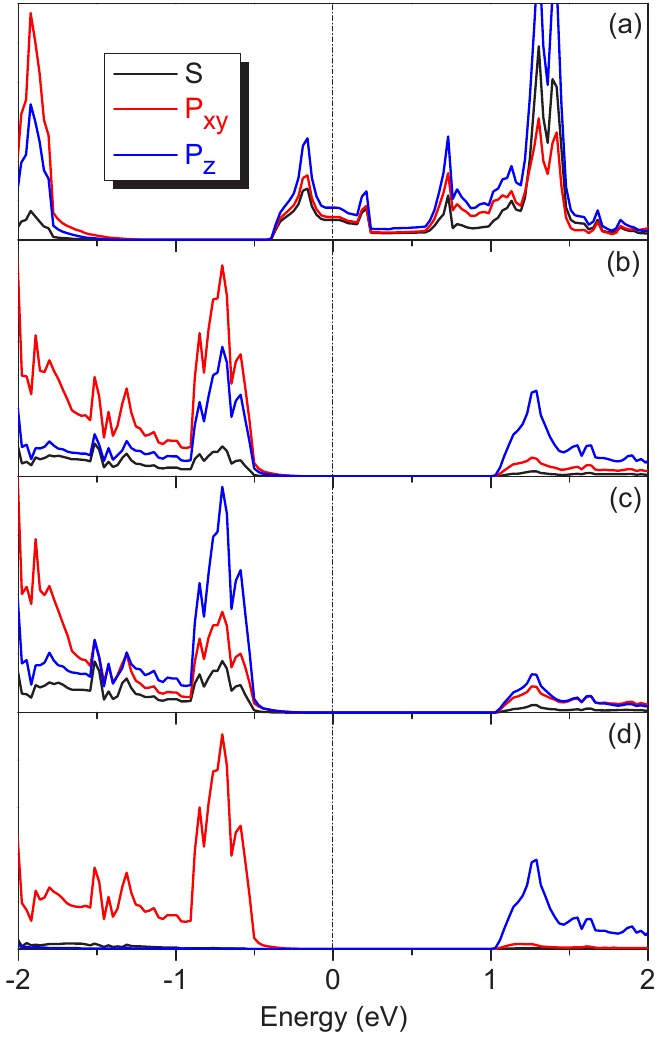}
\caption{%
(Color online) [(a) and (b)] PDOS of the $(111)$-I and $(111)$-I$_{R,(z)}$ structures. [(c) and (d)]PDOS of the $B_{p}$ and $B_{i}$ atoms in $(111)$-I$_{R,(z)}$ structure.}
%\label{fig:eos}
\end{center}
\end{figure}

\begin{figure}[h]
\begin{center}
\includegraphics[width=8.5cm]{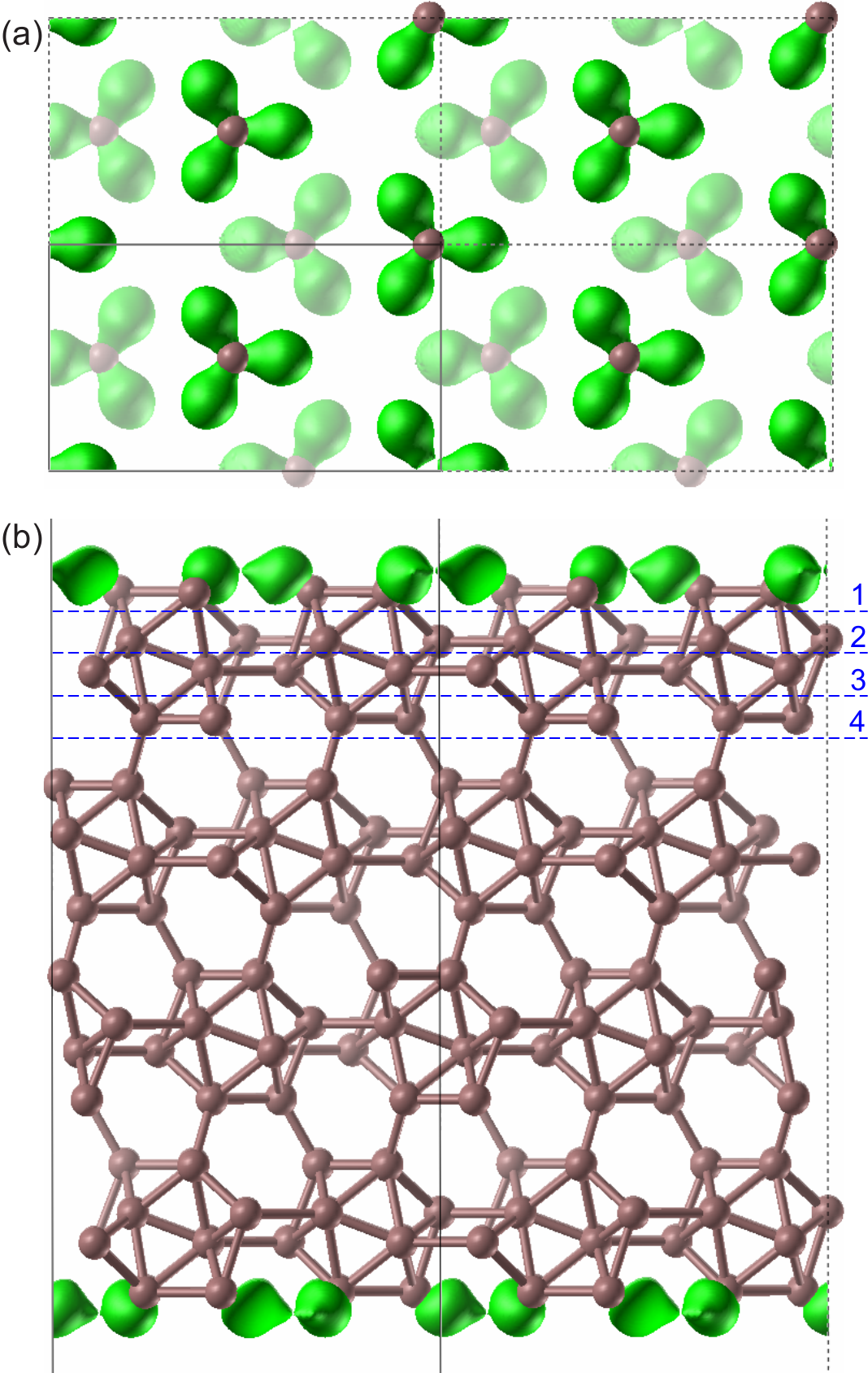}
\caption{%
(Color online) (a) Projection of the charge density difference of the $(111)$-I$_{R,(z)}$ structure along [111] direction. (b) projection of the charge density difference of the $(111)$-I$_{R,(z)}$ structure along $[\bar{1} \bar{1} 2]$ direction. The top four atomic layers are labeled from 1 to 4.}
%\label{fig:eos}
\end{center}
\end{figure}

Symmetry breaking of $B_{12}$ icosahedra results in charge asymmetry on some B-B bonds, and in fact a small degree of ionicity of B-B bonds in the $B_{12}$ icosahedron was predicted in $\alpha$-boron \cite{R03}, while a much greater degree of ionicity was found in the high pressure partially ionic $\gamma$ phase \cite{R04}. It is extremely important and interesting to study the charge transfer/ionicity for the $(111)$-I$_{R,(z)}$ surface. Figure 4 shows the charge density difference between $B_{i}$ and $B_{p}$ atoms \cite{R42}. There is a notable charge transfer from $B_{i}$ to the neighboring $B_{12}$ icosahedra through $B_{p}$ atoms. Bader charges for $B_{i}$ is +0.17$e$, and +0.04$e$ for $B_{p}$ \cite{R43}. Strikingly, the charge transfer keeps the charge state of $B_{p}$ close to its bulk state ($\sim$$+0.05 e$)\cite{R04,R23}, and the significant charge transfer between the $B_{i}$ and $B_{12}$ icosahedra indicates the bridging $B_{i}$-$B_{p}$ bonds are unique polar covalent bonds, which contrasted sharply with the intericosahedral purely covalent $B_{p}$-$B_{p}$ bonds \cite{R03}. Due to the charge transfer, the $B_{i}$-$B_{p}$ bonds (1.793~{\AA}) are much weaker and longer than the $B_{p}$-$B_{p}$ bonds (1.673~{\AA}). In Fig.~4(b), each $B_{12}$ icosahedron comprises 4 atomic layers (labeled from 1 to 4), the charge transfer for these 4 layers should be in the ``$+ - - +$" order with the values of $+0.56, -0.46, -0.46$, and $+0.36 e$, compared with the corresponding values of $+0.20, -0.22, -0.22$, and $+0.24 e$ in $\alpha$-boron. It is clear from these numbers that the surface region as a whole is charge-neutral. The charge transfer of the $(111)$-I$_{R,(z)}$ surface is rebalanced within the top 4 atomic layers (include the $B_{i}$ atoms and the $B_{12}$ icosahedra), which plays an important role in stabilizing the surface.

In conclusion, the most simple and stable reconstructions of the $\alpha$-boron $(111)$ surface has been investigated in detail using \textit{ab initio} evolutionary structure search. Our results show the $(111)$-I$_{R,(z)}$  surface is lower in energy than the earlier reported structures \cite{R26}, and confirm that the classical ECR governs the reconstructions, results in the formation of novel bridging bonds between the $B_{i}$ and $B_{p}$ atoms, leading to the metal-semiconductor transition. In particular, significant charge transfer is responsible for the unique polar-covalent bonds between the $B_{i}$ and $B_{p}$ atoms. Charge redistribution between the $B_{i}$ atoms and the $B_{12}$ icosahedra is one of the key factors stabilizing the reconstruction.

X.F.Z. thanks Philip B. Allen and Maria V. Fernandez-Serra for valuable discussions. This work was supported by the National Science Foundation of China (Grants No. 11174152 and No. 91222111), the National 973 Program of China (Grant No. 2012CB921900), the Program for New Century Excellent Talents in University (Grant No. NCET-12-0278), and the Fundamental Research Funds for the Central Universities (Grant No. 65121009). A.R.O. thanks the National Science Foundation (EAR-1114313, DMR-1231586), DARPA (Grants No. W31P4Q1310005 and No. W31P4Q1210008), DOE (Computational Materials and Chemical Sciences Network (CMCSN) project DE-AC02-98CH10886), CRDF Global (UKE2-7034-KV-11), AFOSR (FA9550-13-C-0037), the Government (No. 14.A12.31.0003) and the Ministry of Education and Science of Russian Federation (Project No. 8512) for financial support, and Foreign Talents Introduction and Academic Exchange Program (No. B08040). Calculations were performed on XSEDE facilities and on the cluster of the Center for Functional Nanomaterials, Brookhaven National Laboratory, which is supported by the DOE-BES under contract no. DE-AC02-98CH10086.

%\newpage

%\newpage


\begin{references}
\bibitem{R42} The charge density difference $\delta\rho = \rho_{AB} - ( \rho_A + \rho_B)$, where $\rho_{AB}$ is the charge density of the $(111)$-I$_{R,(z)}$ surface, $\rho_A$ is the charge density of the $B_{i}$ atoms, and $\rho_B$ is the charge density of the unrelaxed structure which the $B_{i}$ atoms are removed from the $(111)$-I$_{R,(z)}$ surface.
\end{references}
\end{document}